\documentclass[11pt, showpacs,showkeys,superscriptaddress,preprintnumbers,nofootinbib]{revtex4-1}
\usepackage{amsmath,amssymb,bm,mathrsfs}
\usepackage{srcltx}
\usepackage{dsfont}
\usepackage{epsfig}
\usepackage{slashed}
\usepackage{bbold}
\usepackage{psfrag}
\usepackage{xcolor}
\PassOptionsToPackage{caption=false}{subfig}
\usepackage{subfig}
\usepackage{xfrac}
\usepackage{multirow}
\usepackage{booktabs}
\usepackage{physics}
\usepackage[colorlinks=true,linkcolor=blue,citecolor=blue,urlcolor=violet]{hyperref}
\usepackage{cancel}
\usepackage{ulem}
\usepackage[bottom]{footmisc}

\newcommand{\be}{\begin{eqnarray}}
\newcommand{\ee}{\end{eqnarray}}

\def\({\left(}
\def\){\right)}

\definecolor{colorRTD}{rgb}{.2,.2,.7}

\usepackage{soul}

\begin{document}

\title{Forbidden Dark Matter Annihilations into Standard Model Particles} 

\author{Raffaele Tito D'Agnolo}
\email{raffaele-tito.dagnolo@ipht.fr}
\affiliation{Institut de Physique Th\'eorique, Universit\'e Paris Saclay, CEA, F-91191 Gif-sur-Yvette, France} 

\author{Di Liu}
\email{di.liu@desy.de}
\affiliation{Deutsches Elektronen-Synchrotron (DESY), D-22607 Hamburg, Germany}

\author{Joshua T. Ruderman}
\email{ruderman@nyu.edu}
\affiliation{Deutsches Elektronen-Synchrotron (DESY), D-22607 Hamburg, Germany}
\affiliation{
Center for Cosmology and Particle Physics,
Department of Physics, New York University, New York, NY 10003, USA.
}

\author{Po-Jen Wang}
\email{pjw319@nyu.edu}
\affiliation{
Center for Cosmology and Particle Physics,
Department of Physics, New York University, New York, NY 10003, USA.
} 

\begin{abstract}
We present kinematically forbidden dark matter annihilations into Standard Model leptons. This mechanism precisely selects the dark matter mass that gives the observed relic abundance. This is qualitatively different from existing models of thermal dark matter, where fixing the relic density typically leaves open orders of magnitude of viable dark matter masses. Forbidden annihilations require the dark matter to be close in mass to the particles that dominate its annihilation rate. 
We show examples where the dark matter mass is close to the muon mass, the tau mass, or the average of the tau and muon masses.
We find that most of the relevant parameter space can be covered by the next generation of proposed beam-dump experiments and future high-luminosity electron positron colliders. Forbidden dark matter predicts large couplings to the Standard Model that can explain the observed value of $(g-2)_\mu$.
\end{abstract}
\begin{flushright}
DESY 20-225
\end{flushright}
\maketitle

\section{Introduction}

Dark Matter (DM) constitutes more than 80\% of the matter in our Universe, but its microscopic origin is unknown and remains one of the most important open questions in physics.
If DM interactions with Standard Model (SM) particles are responsible for its present abundance, laboratory experiments can shed light on its origin. The best known example  is the Weakly Interacting Massive Particle (WIMP)~\cite{Lee:1977ua, Kolb:1990vq, Gondolo:1990dk, Jungman:1995df}, with abundance set by its electroweak interactions. The WIMP annihilation rate is precisely predicted and testable by current experiments. Well-studied variants allow DM masses in the MeV to GeV range~\cite{Boehm:2003hm,Pospelov:2007mp, Boehm:2003hm,Pospelov:2007mp,Strassler:2006im,ArkaniHamed:2008qn} and predict sharp targets for laboratory experiments~\cite{Izaguirre:2014bca,Izaguirre:2015yja,Essig:2015cda,Alexander:2016aln,Battaglieri:2017aum,Berlin:2018bsc}. Theories of sub-GeV DM have driven a large experimental effort, and are motivating many of the new DM detectors that are being proposed and built today. 

In this paper we present a mechanism that shares the attractive theoretical features of the WIMP and its simplest extensions, but makes two qualitatively distinct predictions: (1) the DM coupling to SM particles can be exponentially larger, and (2) the DM mass is close to the mass of a SM particle, with a relative mass splitting smaller than $\mathcal{O}(1)$. Existing scenarios for thermal DM production, including WIMP-like-variants, predict a relation between the DM coupling and its mass, leaving the DM mass free to vary over several orders of magnitude as the coupling varies.   On the contrary, we identify sharp targets for the DM mass.

We propose a cosmology where DM has all the simple characteristics of a WIMP: a) it is in chemical and kinetic equilibrium with the SM photons when $T \gtrsim m_{\rm DM}$, b) it follows a standard cosmological history, and c) it has a relic density determined by the decoupling of its annihilations into SM particles. As noted in~\cite{DAgnolo:2019zkf}, the first two assumptions are not satisfied by non-thermal DM~\cite{Moroi:1993mb,Dodelson:1993je, Feng:2003xh, Hall:2009bx}, cannibalism~\cite{Carlson:1992fn,Kuflik:2015isi,Pappadopulo:2016pkp,Farina:2016llk}, and late entropy production~\cite{Gelmini:2006pw, Wainwright:2009mq, Berlin:2016vnh}; the third is violated if DM annihilates into hidden sector particles~\cite{Finkbeiner:2007kk,Pospelov:2007mp,Feng:2008ya,Hochberg:2014dra,Hochberg:2014kqa,Evans:2017kti} or has a relic density set by a particle asymmetry~\cite{Kaplan:2009ag,Petraki:2013wwa,Zurek:2013wia}.

In this work we study kinematically forbidden annihilations~\cite{Griest:1990kh, DAgnolo:2015ujb} into light SM particles. When $T\lesssim m_{\rm DM}$, the DM relic density is dominantly depleted by its annihilations into heavier SM particles: $\chi \chi \to \ell_1 \ell_2$, with $m_\chi < (m_{\ell_1}+m_{\ell_2})/2$. The annihilation rate is nonzero at finite temperature.  (So the annihilations are only {\it forbidden} at zero temperature.  Sometimes language isn't perfect: you park on a driveway, drive on a parkway, and are perfectly allowed to annihilate through forbidden channels.)  
If DM has sufficiently large couplings to the SM, there is a small mass window, 
\be
0.3(m_{\ell_1}+m_{\ell_2})\lesssim m_\chi < (m_{\ell_1}+m_{\ell_2})/2, 
\ee
where DM is depleted at the correct rate and can have the observed relic density today.

Kinematically forbidden annihilations of DM were first discussed in~\cite{Griest:1990kh}, where they are referred to as ``forbidden channels." They were subsequently studied as a mechanism for light DM with annihilations to new dark sector particles~\cite{DAgnolo:2015ujb}, and for weak scale DM with annihilations to top quarks~\cite{Delgado:2016umt}.  Here we show that perhaps the simplest incarnation of this mechanism, forbidden annihilations into the SM lepton sector, is still experimentally viable and makes new predictions for the DM mass and annihilation rate that can be tested at future fixed target and collider experiments.  DM could, in principle, experience forbidden annihilations into a variety of light SM particles, but for simplicity we focus on leptons, as a first step. Furthermore, dark sectors coupling to muons can explain a longstanding tension between theory and experiment in the determination of the muon anomalous magnetic moment~\cite{Aoyama:2020ynm, Bennett:2006fi, Hagiwara:2011af, Davier:2010nc}. This tension has recently increased to $4.2\sigma$\footnote{However it should be noted that a recent lattice result~\cite{Borsanyi:2020mff}, not yet included in the theory average, is much closer to the experimental value. due to the measurement} by the $g-2$ collaboration at Fermilab~\cite{Grange:2015fou, Albahri:2021kmg, Albahri:2021ixb, Abi:2021gix}, which is in agreement with the previous measurement at Brookhaven~\cite{Bennett:2006fi}. This discrepancy will be further tested by the J-PARC $g-2$ experiment~\cite{Mibe_2010}. 

As noted above, the most striking phenomenological feature of this mechanism is that it selects a handful of precise values for the DM mass. It would be interesting to build UV models that explain why the DM mass is close to that of a SM particle. A simple possibility would be to consider a twin sector~\cite{Chacko:2005pe, Chacko:2005un, Chacko:2016hvu, Cheng:2018vaj, Craig:2015xla} with a subset of the SM particle content replicated, as was done for instance in~\cite{Craig:2015pha} to make the Higgs mass technically natural. We leave this to future work. The main focus of this paper is to show that these narrow mass windows are selected by experiment.   
This is reminiscent of WIMP DM in supersymmetric extensions of the SM\@. Much of the neutralino parameter space, that  reproduces the observed DM abundance and remains experimentally viable, requires a ``well-tempered" mixture of neutralinos  with nearly degenerate masses~\cite{ArkaniHamed:2006mb}. Analogously, for sub-GeV DM annihilating to SM particles, we find that only the narrow mass window, allowing for forbidden annihilations, can at the same time give the observed relic density and naturally evade CMB constraints on energy injection at the recombination epoch~\cite{Padmanabhan:2005es, Aghanim:2018eyx}.

The rest of the paper is organized as follows.  In Section~\ref{sec:relic} we review the calculation of the relic density with kinematically forbidden annihilations, and we derive the exponential factor that characterizes Forbidden DM phenomenology.  In Section~\ref{sec:models} we show how to construct explicit models with DM annihilating to SM leptons via a scalar mediator and study their phenomenology. In Section~\ref{sec:vector} we perform the same exercise for vector mediators. We conclude in Section~\ref{sec:conclusion}.

\section{Forbidden Annihilations}\label{sec:relic} 
The discussion in this Section mostly follows~\cite{DAgnolo:2015ujb}.
Consider a DM particle $\chi$ dominantly annihilating into two SM particles $\ell_{1,2}$ with mass $m_\chi < (m_{\ell_1}+m_{\ell_2})/2$. Early on ($T\gg m_\chi$) DM is in thermal equilibrium with the SM photons and the $\chi$ Boltzmann equation is
\be
\dot n_\chi + 3 H n_\chi = -\langle \sigma_\chi v \rangle n_\chi^2 + \langle \sigma_\ell v \rangle n_{\ell_1}^{\rm eq} n_{\ell_2}^{\rm eq}\, , \label{eq:BE}
\ee
where $\sigma_\chi$ is the annihilation cross section, $\sigma_\chi \equiv \sigma(\chi \chi \to \ell_1 \ell_2)$, while $\sigma_\ell$ is the cross section for the inverse process, which is kinematically allowed also at zero temperature. In equilibrium the right-hand side of Eq.~\eqref{eq:BE} vanishes and the thermally averaged forbidden annihilation rate is
\be
\langle \sigma_\chi v \rangle= \langle \sigma_\ell v \rangle \frac{n_{\ell_1}^{\rm eq} n_{\ell_2}^{\rm eq}}{(n_\chi^{\rm eq})^2} \simeq \langle \sigma_\ell v \rangle e^{- 2\Delta x}\, ,
\ee 
where $x\equiv m_\chi/T$ and $\Delta\equiv (m_{\ell_1}+m_{\ell_2}-2 m_\chi)/2 m_\chi$. In the following we never consider $\Delta$ large enough to cross the threshold beyond which $3\to 2$ annihilations dominate the abundance~\cite{Cline:2017tka}. 

When the annihilation rate becomes slower than the Hubble expansion, DM is no longer in equilibrium with the SM thermal bath.  The DM relic density stops being depleted and is related to its value today by a simple redshift factor. DM leaves equilibrium when $\chi$ and $\ell_{1,2}$ are non-relativistic. So we can expand the thermally-averaged cross section for the inverse annihilations
\be
\langle \sigma_\ell v \rangle \simeq \bar\sigma_{\ell,s} +(3/2x) \bar\sigma_{\ell,p}+\mathcal{O}(1/x^2). 
\ee

To estimate the relic density we can use a sudden freeze-out approximation, i.e.~we assume that DM leaves equilibrium abruptly when $\langle \sigma_\chi v \rangle n_\chi^{\rm eq} \approx H$. 
This approximation defines the freeze-out temperature
\be
(1+2 \Delta)x_f &\simeq& 21+\log\left[\frac{g_{\ell_1}g_{\ell_2}}{g_\chi\sqrt{g_*}}\frac{m_\chi \bar\sigma_{\ell,s}}{1.1\;{\rm pb\times GeV}}\right]+ \log[(2 r(1+\Delta)-r^2)^{\frac{3}{2}} \sqrt{x_f}]\, ,  \label{eq:FO}
\ee
where $g_{\ell_{1,2}}, g_\chi$ count internal degrees of freedom, $g_*=g_*(x_f)$ counts the relativistic degrees of freedom in the Universe, $r\equiv m_{\ell_2}/m_\chi$, and we have assumed an $s$-wave inverse annihilation process $ \langle \sigma_\ell v \rangle \simeq \bar\sigma_{\ell,s}$.

Using Eq.~\eqref{eq:FO} we obtain an estimate of today's relic density
\be
\frac{\Omega_\chi}{\Omega_{\rm DM}} \simeq \left(\frac{0.3\;{\rm pb}}{\bar \sigma_{\ell, s}e^{-2 \Delta x_f}}\right)\frac{g_\chi^2 \sqrt{g_*}x_f}{g_{\ell_1}g_{\ell_2}g_{*S}(r (2 - r + 2 \Delta))^{\frac{3}{2}}} \, .
\ee
Given that $x_f \sim 10 - 25$ for $m_\chi \sim {\rm keV} - {\rm TeV}$~\cite{DAgnolo:2015ujb}, the exponential enhancement of the relic density can be sizable for $\Delta \sim \mathcal{O}(1)$, leading to very different predictions for laboratory experiments compared to kinematically allowed annihilations ($e^{-2 \Delta x_f}\to1$). 

Taking $\Delta \gg 1$ at fixed couplings, reduces exponentially the annihilation rate, rapidly leading to too large of a DM abundance. Therefore the mass window where this mechanism can give the observed relic abundance is limited, as we show in the following.

A second key feature of this mechanism is that kinematically forbidden DM annihilations naturally evade the stringent CMB bounds on energy injection at the recombination epoch~\cite{DAgnolo:2015ujb}. A WIMP-like thermal relic with $s$-wave annihilations is currently excluded below $m_\chi\lesssim 10$~GeV by Planck~\cite{Aghanim:2018eyx}. However the Boltzmann suppression of the rate at $T\lesssim$~eV, for Forbidden DM, insures that sub-GeV thermal relics are consistent with experiment, making annihilations to SM leptons with DM masses $m_\chi \ll 10$~GeV still viable. 

\section{Annihilations to SM Leptons}\label{sec:models}
After reviewing the general features of Forbidden DM annihilations, we are now in a position to study annihilations into SM leptons. Compared to annihilations to dark sector particles~\cite{DAgnolo:2015ujb} our models make the qualitative different prediction that DM must have a mass very close to that of a SM particle. To make these statements more concrete we construct simple models and systematically discuss all phenomenologically distinct possibilities for annihilations into SM leptons.

We start by discussing scalar interactions of DM\@. A vector mediator requires additional model building to cancel anomalies and we treat this possibility in the next Section. 
In this Section DM is a Dirac fermion $\chi$ that couples to the SM through the scalar mediator $\phi$. Below the EW breaking scale we can write the effective Lagrangian

\be
-\mathcal{L}\supset g_{V}^{ij} \phi \bar e_i e_j+g_{A}^{ij} \phi \bar e_i \gamma_5 e_j + g_{V\chi} \phi \bar \chi \chi + g_{A\chi} \phi \bar \chi \gamma_5 \chi . \label{eq:scalar}
\ee
 
Inverse annihilations are $s$-wave if $g_{A\chi}\neq 0$ and $p$-wave if $g_{A\chi}$ vanishes. If $g_{A}^{ij}=0$, the $s$-wave component of the cross section is suppressed by the DM-lepton mass difference. At leading order in $v$ for $m_{\ell_1}=m_{\ell_2}=m_{\ell}$ we have 
\be
\langle \sigma_\ell v \rangle = \frac{8\pi \alpha_A\left[m_\ell^2\left(\alpha_{\chi A}+\alpha_{\chi V}\right)-m_\chi^2 \alpha_{\chi V}\right]}{(m_\phi^2-4 m_\ell^2 )^2+m_\phi^2 \Gamma_\phi^2}\sqrt{1-\frac{m_\chi^2}{m_\ell^2}}+\mathcal{O}(v^2)\, .
\ee

Above the scale of EW symmetry breaking Eq.~\eqref{eq:scalar} can be UV completed to
\be
-\mathcal{L}_{\rm EW}\supset  y^{ij} \frac{\phi}{M} \bar L_i H e_{R, j}+{\rm h.c.} \, , \label{eq:scalarEW}
\ee
so that $g_{V}^{ij} \sim \Re[y^{ij}] v/M$ and $g_{A}^{ij} \sim i \Im[y^{ij}] v/M$. At the scale $M$ we can generate Eq.~\eqref{eq:scalarEW} by integrating out vector-like leptons $X_{L,R}$ with Yukawa couplings to the SM, 
\be
-\mathcal{L}_{\rm UV}\supset \lambda_i \bar L_i H X_R + M \overline X_L X_R + \lambda_j \phi \overline X_L e_{R, j}+{\rm h.c.}
\ee
At colliders $X_{L, R}$ are produced in pairs through their coupling to hypercharge. Even for the smallest Yukawa couplings that we consider in the following, $\lambda_i \lambda_j \simeq 10^{-4}$, the new leptons decay promptly to a SM lepton and a massive SM boson: $W, Z, h$.  For simplicity we only consider $M\gtrsim {\rm TeV}$, which is easily consistent with LHC and LEP constraints~\cite{Sirunyan:2019ofn, Sirunyan:2019bgz, Aad:2015dha, Holdom:2014rsa, Achard:2001qw}. With these masses, $X_{L,R}$ do not appreciably affect the calculation of the relic density.  
For simplicity, we consider a single annihilation channel at a time, for instance $y^{ij}=0$ for $i,j\neq e$. This allows us to systematically explore all the relevant DM phenomenology.  At the end of this Section, we also discuss a light leptophilic Higgs-like mediator: $y^{ij}=\delta^{ij} y_j$, $y_i/y_j=m_{\ell_i}/m_{\ell_j}$. However, in the viable parameter space, this does not qualitatively differ from annihilations into muon pairs. We discuss flavor constraints separately for each $y^{ij}$ flavor structure.

\subsection{$\chi\chi \to e^+e^-$}\label{sec:ele}
Forbidden annihilations into $e^+e^-$ require $m_\chi \sim m_e$.
Annihilations into electron-positron pairs decouple from the thermal bath when $T\sim m_e/10$. As a consequence, DM is in thermal equilibrium during BBN, giving an unacceptably large contribution to the expansion rate of the Universe, which alters the primordial element abundances measured today. Moreover, CMB measurements set a constraint on DM entropy transferred into electrons after neutrinos decouple~\cite{Boehm:2013jpa}. The combined constraint on our Dirac DM particle is $m_\chi>10.9$ MeV~\cite{Sabti:2019mhn}. Even if DM were a real scalar we would have a comparable bound: $m_\chi>6.4$ MeV~\cite{Sabti:2019mhn}. These bounds conservatively apply to DM only, if the mediator is light it will also contribute to the expansion rate, leading to stronger bounds.

 In summary cosmological constraints make forbidden annihilations into electrons $m_\chi \lesssim m_e$ unfeasible experimentally.

 \subsection{$\chi\chi \to e^{\pm}\ell^{\mp}$}
The annihilation channels with $\ell =\mu, \tau$ are not viable due to tree-level flavor constraints. When $m_\chi< (m_\ell - m_e)/2$, the decay width
\be
\Gamma(\ell \to e \chi \chi)\simeq \frac{(\alpha_V+\alpha_A)\alpha_{\chi V}}{2\pi}\frac{m_\ell^5}{(m_\phi^2-m_\ell^2)^2}\left(\frac{\Delta}{1+\Delta}\right)^{7/2} \label{eq:width}
\ee
is always larger than the experimental uncertainty on $\Gamma(\ell \to e \bar \nu \nu)$~\cite{Tanabashi:2018oca}, for couplings compatible with Forbidden DM annihilations. In Eq.~\eqref{eq:width} $\alpha_{V, A}\equiv g_{V, A}^2/4\pi$, with $g$ defined in Eq.~\eqref{eq:scalar}.

\subsection{$\chi\chi \to\mu^+\mu^-$}\label{sec:mumu}
Annihilations to $\mu^+\mu^-$ are experimentally viable  and have interesting phenomenology. A DM particle predominantly annihilating to muon pairs with mass
\be
0.9 m_\mu \lesssim m_\chi \lesssim m_\mu
\ee
and $\mathcal{O}(1)$ couplings, can have the observed relic density with mediator couplings to the SM as large as $\alpha_{\mu\mu}^{\text{max}}=10^{-7}$. This is about three orders of magnitude larger than the prediction from ordinary annihilations with the same $\mathcal{O}(1)$ DM couplings to the mediator. This dark sector can also explain the current discrepancy between the measurement  of the muon anomalous magnetic moment and its SM prediction~\cite{Bennett:2006fi, Hagiwara:2011af, Davier:2010nc}. Alternative explanations from DM models include~\cite{Agrawal:2014ufa, Kahn:2018cqs, Mohlabeng:2019vrz, Krnjaic:2019rsv}.

\begin{figure*}[t]
\includegraphics[width=0.9\textwidth]{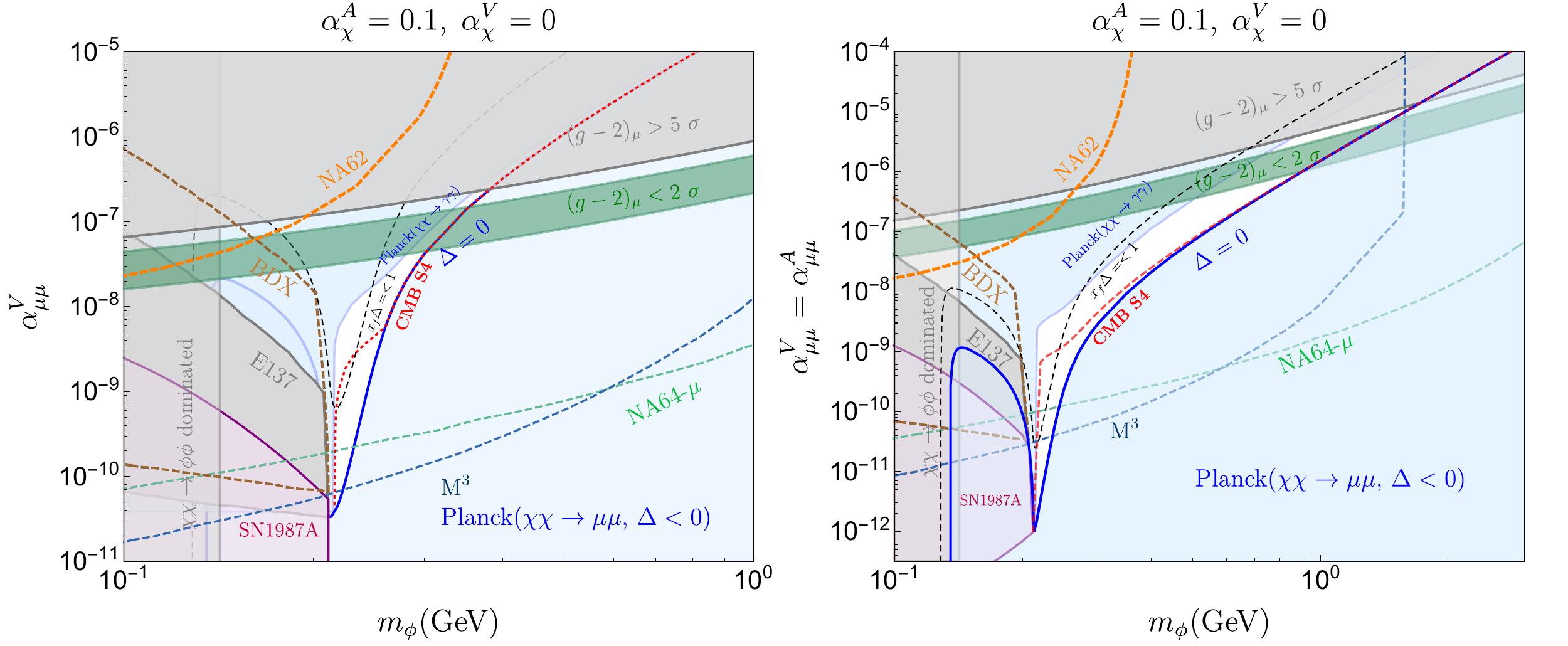}
\includegraphics[width=0.9\textwidth]{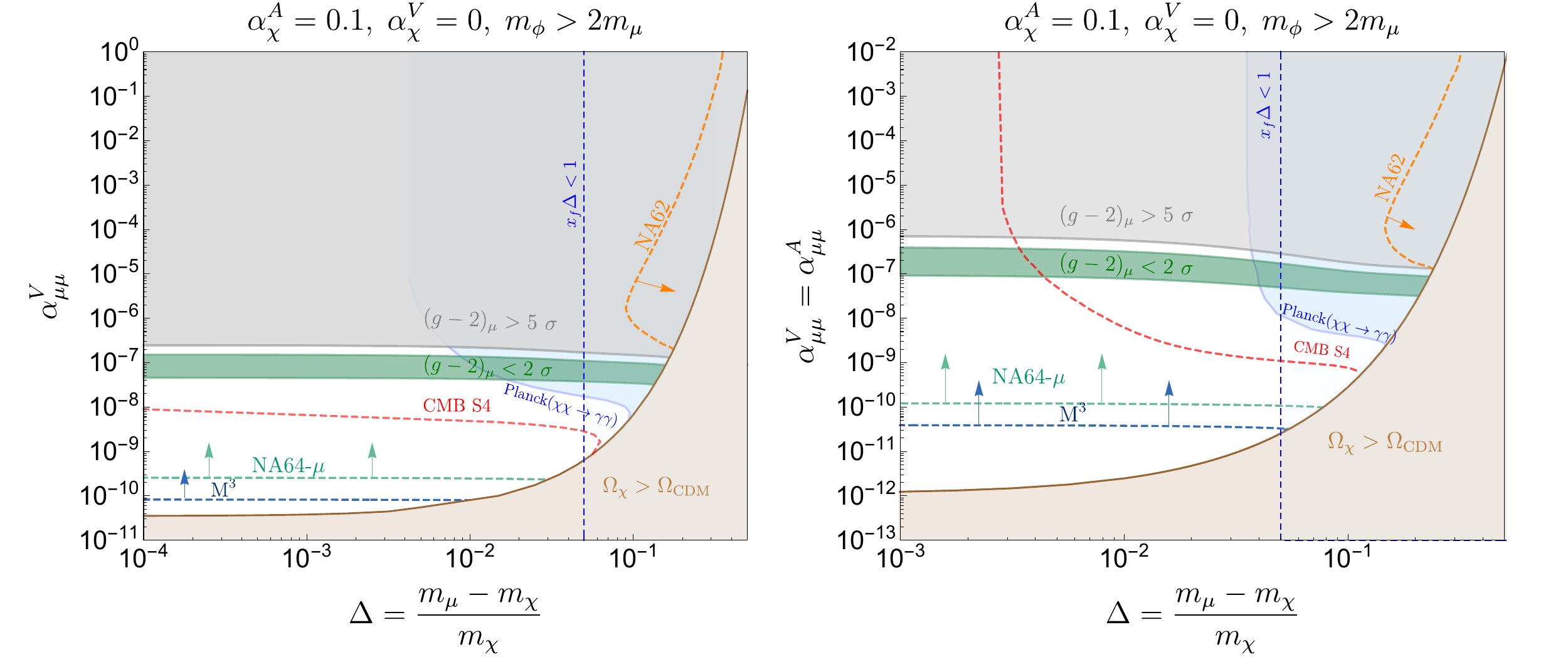}
\caption{Theoretical and experimental constraints on Forbidden DM annihilations into $\mu^+\mu^-$. We show constraints from the E137 electron beam-dump experiment~\cite{Marsicano:2018vin}, excess cooling of SN1987A~\cite{Hirata:1987hu}, measurements of the muon anomalous magnetic moment~\cite{Bennett:2006fi, Hagiwara:2011af, Davier:2010nc}, and Planck bounds on annihilations at the recombination epoch~\cite{Aghanim:2018eyx}. We include projections for CMB-S4 \cite{Abazajian:2016yjj} and future beam-dump experiments from BDX~\cite{Bondi:2017gul}, ${\rm M}^3$~\cite{Berlin:2018bsc, Kahn:2018cqs}, NA62~\cite{Krnjaic:2019rsv, NA62:2017rwk}, and NA64-$\mu$~\cite{Chen:2018vkr, Gninenko:2014pea, Gninenko:2001hx}. In the upper (lower) row, we choose $\Delta$ ($m_\phi$) to match the observed relic density in the white region. Ordinary annihilations into pairs of mediators set the relic density in the shaded gray region.  In the brown area, Forbidden DM has a larger abundance than the observed DM abundance.
}
\label{fig:mumu}
\end{figure*}

In Fig.~\ref{fig:mumu} we show the viable parameter space for forbidden annihilations (in white) and its overlap with the region where $(g-2)_\mu$ is within $2\sigma$ of its measurement (in green). The contribution to $(g-2)_\mu$ from a scalar mediator coupled to a muon and another lepton $\ell$ is~\cite{Lindner:2016bgg, Kowalska:2017iqv}
\be
(g-2)_\mu= \frac{\lambda^2_\phi}{\pi} \int_0^1 dx \frac{\alpha_V^{\mu \ell} P_1^+ (x)+\alpha_A^{\mu \ell} P_1^- (x)}{(1-x)(1-\lambda_\phi^2 x)+\epsilon_\ell^2 \lambda_\phi^2 x} \label{eq:g2}
\ee
where $P_1^\pm =x^2 (1-x \pm \epsilon_\ell)$, $\epsilon_\ell=m_\ell/m_\mu$, and $\lambda_\phi = m_\mu/m_\phi$. 

The most interesting qualitative features of Fig.~\ref{fig:mumu} are related to the calculation of the relic density. The two upper plots show the parameter space in the $\alpha_{\mu\mu}-m_\phi$ plane.  In the white region, $\Delta$ is chosen at each point to match the observed relic density. The most interesting aspect of the Figure is the blue shaded area corresponding to the Planck bound on late time DM annihilations. On the boundary of the constraint $\Delta=(m_\mu-m_\chi)/m_\chi=0$ and in the excluded region DM is WIMP-like ($\Delta <0$), with its relic density set by ordinary annihilations ($m_\chi >m_\mu$). The allowed parameter space is still open thanks to the suppression of the late time annihilation rate in the forbidden regime. When $\Delta>0$, late time $\bar{\chi}\chi\rightarrow \mu^+\mu^-$ annihilations are exponentially suppressed $\langle \sigma_\mu v \rangle \sim e^{-\Delta m_\chi/T}$. 
On the contrary, $\bar{\chi}\chi\rightarrow \gamma\gamma$ annihilations via a muon loop are negligible at freezeout, but can affect the CMB at an observable level.  We show in blue the corresponding constraint from Planck \cite{Slatyer:2015jla, Aghanim:2018eyx} and in red a projection from CMB-S4 measurements~\cite{Abazajian:2016yjj}. 

A second interesting feature is 
that we can be deep in the forbidden regime ($\Delta \sim \mathcal{O}(1)$) only if annihilations are almost resonant $2 m_\chi \sim m_\phi$. This is the only way to overcome the Boltzmann suppression of the rate while maintaining all couplings perturbative. However this can not be done indefinitely if we fix the mediator-DM coupling, due to another feature of forbidden annihilations. The denominator of the relevant annihilation cross section is $\langle \sigma(\bar \chi \chi \to \mu^+\mu^-) v \rangle \sim 1/[(m_\phi^2-4m_\mu^2)^2+\Gamma_\phi^2 m_\phi^2]$. If we take $m_\phi^2=4m_\mu^2$, $\phi$ can decay to a pair of DM particles, giving an irreducible contribution to the width $\Gamma_\phi$ independent of the $\phi$ coupling to the SM\@. To reduce the width of $\phi$ we can take $m_\phi < 2 m_\mu$. However in this regime $(m_\phi^2-4m_\mu^2)^2\simeq (-2 \Delta+\Delta^2)^2$. So at some point increasing $\Delta$ leads to too much DM, as shown by the brown area in Fig.~\ref{fig:mumu}. The only exception to this argument is when $m_\chi=m_\mu$. In this case we can keep decreasing the coupling to the SM and the mediator width way past our threshold and still obtain the observed relic density~\cite{Feng:2017drg}.

The last phenomenological highlight in the plots, visible in all four panels of Fig.~\ref{fig:mumu}, is that the vast majority of viable parameter space can be probed experimentally in the near future, at NA64-$\mu$~\cite{Chen:2018vkr} and ${\rm M}^3$~\cite{Berlin:2018bsc, Kahn:2018cqs}. 

The two columns of Fig.~\ref{fig:mumu} differ in the choice of coupling to muons. In the left panel the pseudoscalar coupling is turned off, while in the right panel it is set equal to the scalar coupling. This does not strongly affect the phenomenology. In the two lower panels we show the $\alpha_{\mu\mu}-\Delta$ plane. At each point of the white region, $m_\phi$ is chosen to match the observed relic density. The two panels show that most  constraints are on the mediator $\phi$. An important exception is energy injection into the CMB\@. When $m_\phi< 2 m_\mu$, (not shown in the plot) the $\bar{\chi}\chi\rightarrow \gamma\gamma$ annihilation cross section is strongly enhanced if $m_\phi \simeq 2 m_\chi$, since $\sigma(\bar{\chi}\chi\rightarrow \gamma\gamma)\sim1/[(m_\phi^2-4m_\chi^2)^2+m_\phi^2\Gamma_\phi^2]$. When $m_\phi \simeq 2 m_\chi$ the $\gamma\gamma$ cross section is proportional to $1/\Gamma(\phi \to {\rm SM})^2$, while forbidden annihilations that set the relic density are not resonant, $\langle \sigma_\mu  v\rangle \sim 1/[(m_\phi^2-4m_\mu^2)^2+\Gamma_\phi^2 m_\phi^2]$.
Constraints from E137, from the relic density calculation, and from the CMB exclude the entire forbidden parameter space in this region. 
In the opposite limit (shown in the plot), $m_\phi> 2 m_\mu$, there is sizable allowed parameter space.

In Fig.~\ref{fig:mumu} we also display, as shaded areas, constraints from the E137 electron beam-dump experiment~\cite{Marsicano:2018vin} and excess energy loss of SN1987A~\cite{Hirata:1987hu}.  We include projections from future beam-dump experiments: BDX~\cite{Bondi:2017gul}, LDMX~\cite{Berlin:2018bsc, Kahn:2018cqs},  ${\rm M}^3$~\cite{Berlin:2018bsc, Kahn:2018cqs}, NA62~\cite{Krnjaic:2019rsv, NA62:2017rwk}, and NA64-$\mu$~\cite{Chen:2018vkr,Gninenko:2014pea, Gninenko:2001hx}. When $m_\phi \lesssim m_\chi$ annihilations into pairs of mediators become the dominant source of depletion of the relic density, and DM has a larger abundance than the observed DM abundance for the chosen $\chi-\phi$ coupling.

The experiments E137 and BDX search for signals at electron beam dumps with  $10^{20}$ and $10^{22}$ electrons-on-target, respectively. The constraints arise from direct production of the mediator $\phi$ via $\mu+N\rightarrow \mu+N+\phi$, where $N$ is a nucleon. $M^3$, NA64$-\mu$, and NA62 are missing momentum experiments. The first two exploit the same production mechanism ($\mu N\rightarrow\mu N\phi$) using muon beams at 15 GeV(175 GeV) for $M^3$ (NA64$-\mu$). NA62 searches for $K^+\rightarrow \mu\nu\phi (\phi\rightarrow \text{invisible})$ from Kaons produced using the CERN SPS proton beam. Finally the SN1987A constraint arises from excess energy loss from $\phi$ production through the Primakoff effect $\gamma N\rightarrow N \phi$. 

\subsection{$\chi\chi \to \mu^\pm\tau^\mp$}
\begin{figure*}[t]
\includegraphics[width=0.9\textwidth]{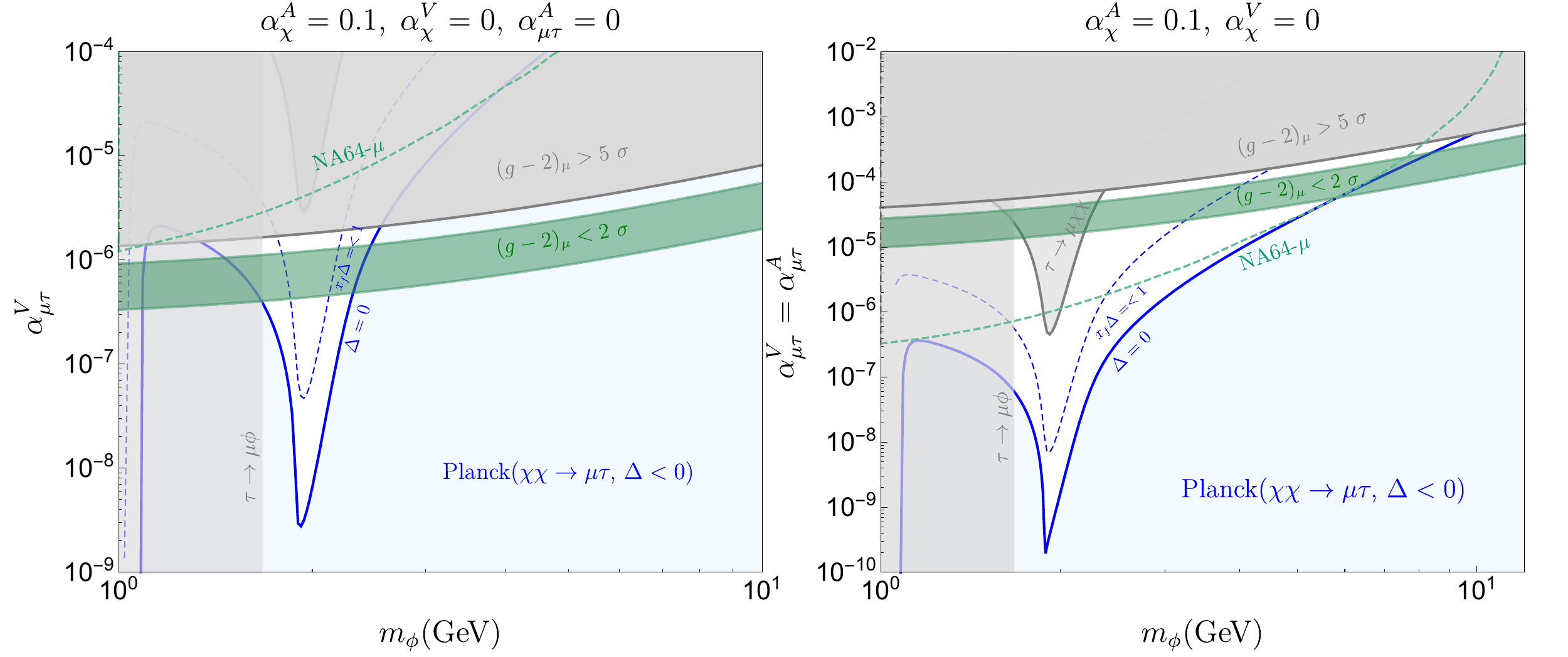}
\includegraphics[width=0.9\textwidth]{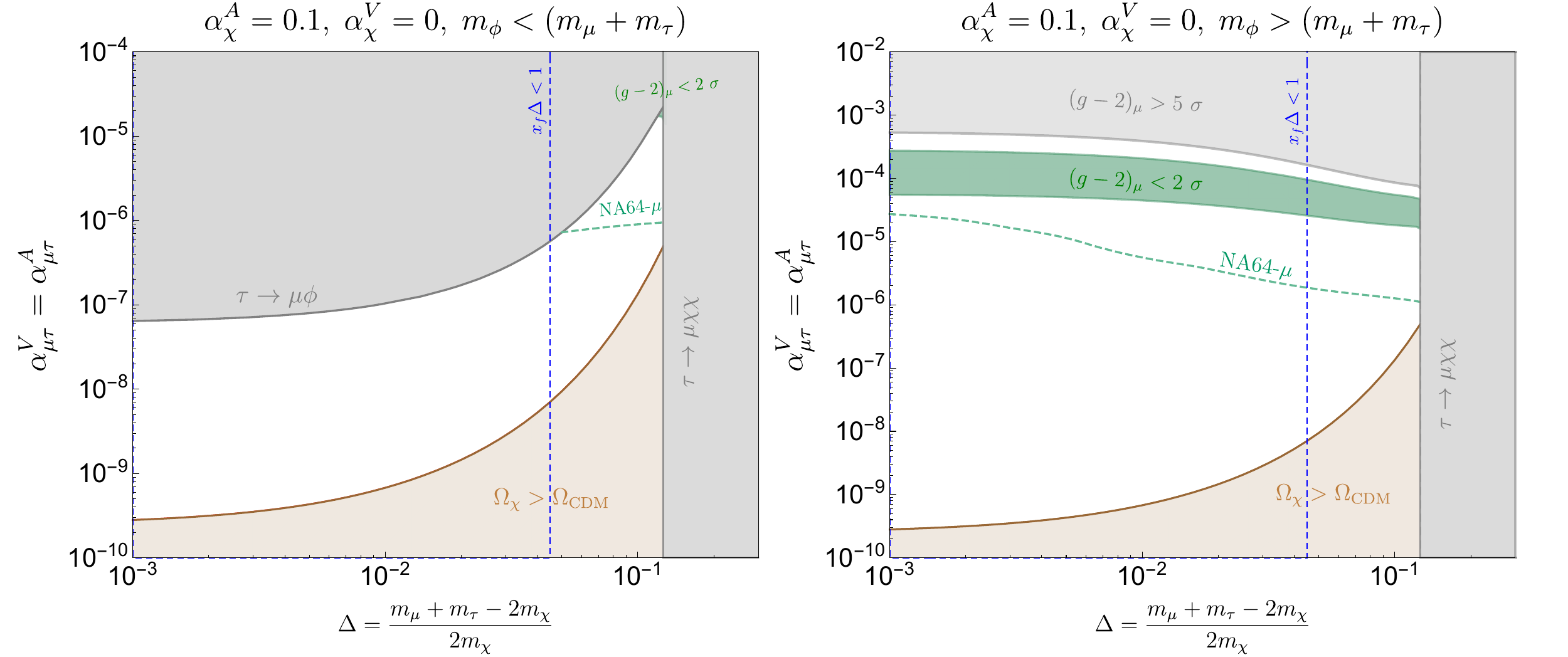}
\caption{
Theoretical and experimental constraints on Forbidden DM annihilations into $\tau+\mu$. We show constraints from measurements of the $\tau$ width $\Gamma(\tau\to \mu \bar \nu_\mu \nu_\tau)$~\cite{Tanabashi:2018oca}, measurements of the muon anomalous magnetic moment~\cite{Bennett:2006fi, Hagiwara:2011af, Davier:2010nc}, and Planck constraints on DM annihilations~\cite{Aghanim:2018eyx}. We include a  projection from the future beam-dump experiment NA64-$\mu$~\cite{Chen:2018vkr}. Due to its flavor violating coupling DM is produced at negligible rates at other beam dump experiments listed in the caption of Fig.~\ref{fig:mumu}. In the upper (lower) row, we choose $\Delta$ ($m_\phi$) to match the observed relic density in the white region. We shade in gray the region where ordinary annihilations into pair of mediators set the relic density. In the brown area Forbidden DM has too large of an abundance.
}
\label{fig:mutau}
\end{figure*}

The results for annihilations to $\mu\tau$ pairs are displayed in Fig.~\ref{fig:mutau}. The main phenomenological features, including the correlation between large $\Delta$ and annihilations on resonance and the forbidden relaxation of the CMB bound on DM annihilations, are unchanged compared to $\chi\chi \to \mu^+\mu^-$ and were discussed in the previous Section. The main qualitative difference is that production through $\tau$'s or through $\mu$'s, followed by an off-shell decay to a $\tau$, is sufficiently suppressed at beam-dump experiments not to yield any constraint. For the same reason there are no proposed experiments that can probe the whole parameter space in Fig.~\ref{fig:mutau}. Nonetheless NA64-$\mu$~\cite{Chen:2018vkr,Gninenko:2014pea, Gninenko:2001hx} retains some sensitivity, since its center of mass energy is sufficient for the process $\mu N \to \tau N \phi$ to take place.
 
The second novelty compared to $\chi\chi \to \mu^+\mu^-$ is that flavor constraints play an important role. The precise measurement of the $\tau$ width $\Gamma(\tau\to \mu \bar \nu_\mu \nu_\tau)$~\cite{Tanabashi:2018oca} forces $m_\chi > (m_\tau-m_\mu)/2$. Otherwise the branching ratio $\tau \to \mu \chi \chi$ is unacceptably large in the ``forbidden" parameter space. This still leaves open the window
\be
 (m_\tau-m_\mu)/2 < m_\chi \lesssim (m_\tau+m_\mu)/2
\ee

for DM annihilations. The new leptophilic force mediated by $\phi$ can ease the tension in the measurement of $(g-2)_\mu$, as shown in Fig.~\ref{fig:mutau} and in Eq.~\eqref{eq:g2}. 

\subsection{$\chi \chi \to \tau^+\tau^-$}
\begin{figure*}[t]
\includegraphics[width=0.9\textwidth]{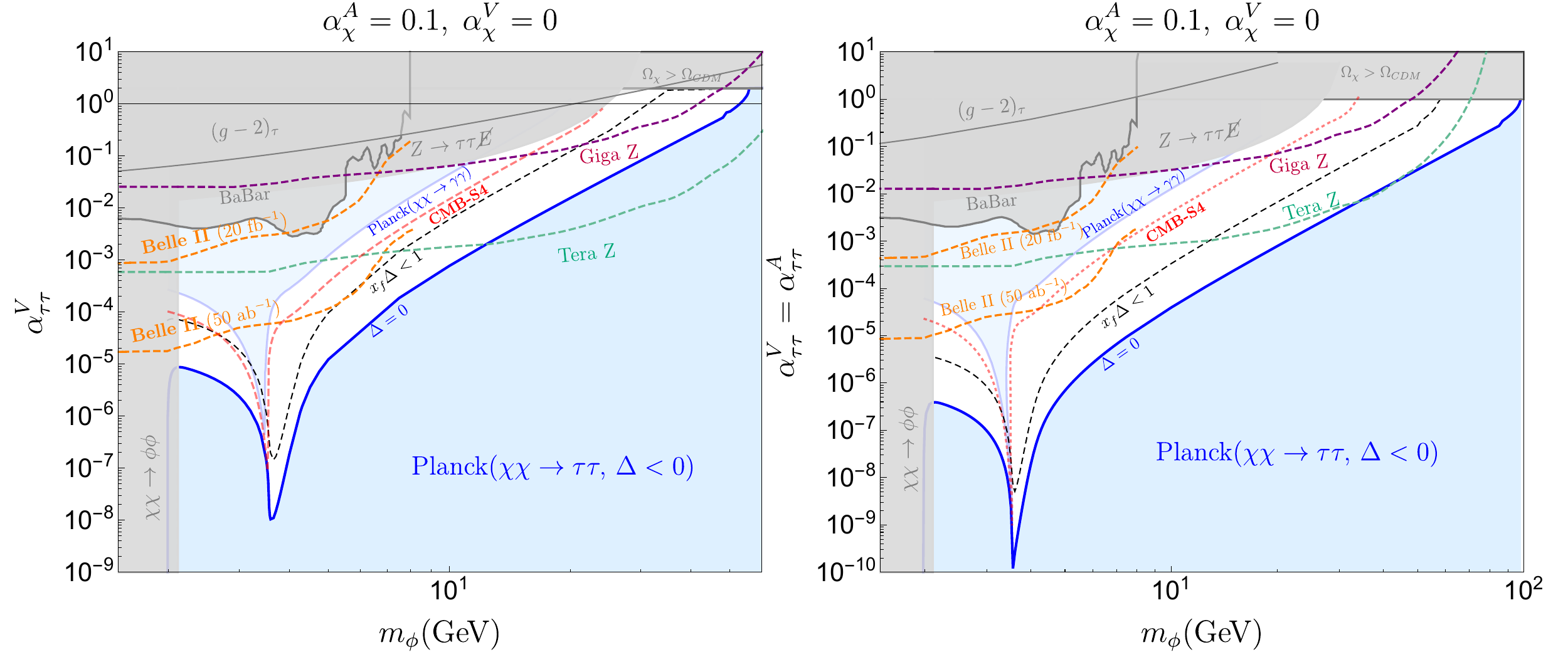}
\includegraphics[width=0.9\textwidth]{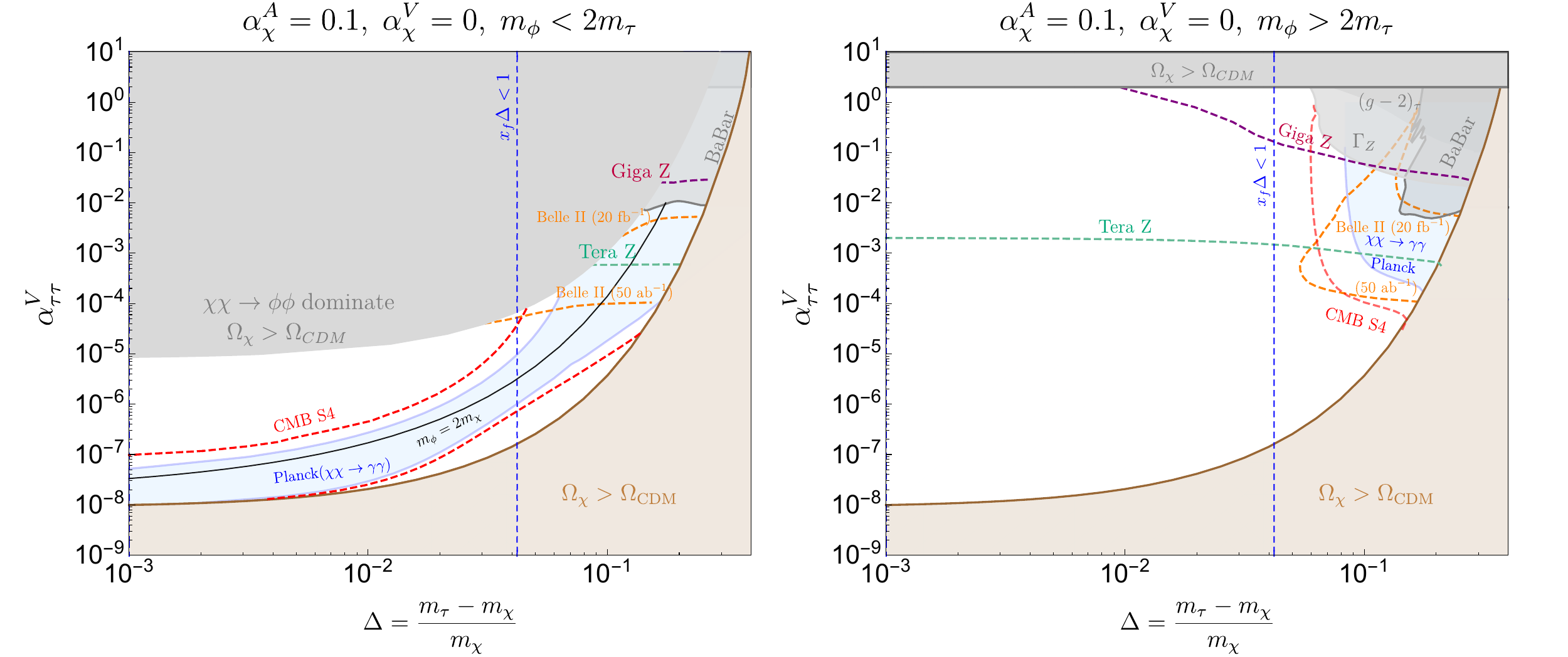}
\caption{
Theoretical and experimental constraints on Forbidden DM annihilations into $\tau^+\tau^-$. We show constraints from searches at BaBar for $e^+e^-\to \phi \gamma$~\cite{Dolan:2017osp, Lees:2017lec, Chen:2018vkr}, searches at LEP for $Z\to \bar \tau \tau$+MET~\cite{Chen:2018vkr, Tanabashi:2018oca}, the LEP measurement of the $Z$ width~\cite{ALEPH:2005ab}, the tau anomalous magnetic moment~\cite{Beresford:2019gww}, and Planck constraints on DM annihilations~\cite{Aghanim:2018eyx}. We include projections for CMB-S4 \cite{Abazajian:2016yjj},  Belle II~\cite{Dolan:2017osp,Abe:2010gxa} and future $Z$-factories~\cite{Liu:2017zdh, Gomez-Ceballos:2013zzn, dEnterria:2016fpc, dEnterria:2016sca, CEPC-SPPCStudyGroup:2015csa}. In the upper (lower) row, we choose $\Delta$ ($m_\phi$) to match the observed relic density in the white region. We shade in gray the region where ordinary annihilations to a pair of mediators set the relic density. In the brown area Forbidden DM has too large of an abundance. 
}
\label{fig:tautau}
\end{figure*}

Annihilations to $\tau^+\tau^-$ are experimentally viable and share a number of qualitative features with annihilations to $\mu^+\mu^-$. Also in this case forbidden annihilations are crucial to comply with the CMB bound on energy injection~\cite{Aghanim:2018eyx}, as shown in the upper panels of Fig.~\ref{fig:tautau}. Furthermore, large Boltzmann suppressions require annihilations close to resonance, as shown in the lower panels of Fig.~\ref{fig:tautau}, but at large $\Delta$ we can not avoid having too large of a DM abundance. The viable DM masses are
\be
0.8 m_\tau \lesssim m_\chi \lesssim m_\tau
\ee
and we can have couplings to the SM as large as $\alpha_{\tau\tau}^{\rm max}\simeq 1$. Experimental constraints are rather different from the $\mu^+\mu^-$ case. First of all DM and the new mediator do not contribute significantly to $(g-2)_\mu$, but do affect the tau anomalous magnetic moment~\cite{Beresford:2019gww} and we show the corresponding $2\sigma$ constraint in Fig.~\ref{fig:tautau}. Secondly, the main constraints arise from $e^+e^-$ circular colliders, as depicted in Fig.~\ref{fig:tautau}. We show constraints from searches at BaBar for $e^+e^-\to \phi \gamma$~\cite{Dolan:2017osp, Lees:2017lec, Chen:2018vkr}, searches at LEP for $Z\to \bar \tau \tau$+MET~\cite{Chen:2018vkr, Tanabashi:2018oca}, and the LEP measurement of the $Z$ width~\cite{ALEPH:2005ab}. We also show the projected sensitivity on the same processes from Belle II~\cite{Dolan:2017osp,Abe:2010gxa} and future $Z$-factories~\cite{Liu:2017zdh, Gomez-Ceballos:2013zzn, dEnterria:2016fpc, dEnterria:2016sca, CEPC-SPPCStudyGroup:2015csa}. The main point in common with $\mu^+\mu^-$ annihilations is the CMB constraint on energy injection from $\bar{\chi}\chi\rightarrow\gamma\gamma$. The parameter space at $m_\phi < 2m_\tau$ is not completely excluded, as shown in the bottom-left panel of Fig.~\ref{fig:tautau}, but it is significantly more constrained than the one at $m_\phi > 2 m_\tau$, due to the resonant effect described in the $\mu^+\mu^-$ case. When $m_\phi \simeq 2m_\chi$, $\sigma(\bar{\chi}\chi\rightarrow\gamma\gamma)$ is enhanced because $1/[(m_\phi^2-4m_\chi^2)^2+m_\phi^2 \Gamma_\phi^2] \sim 1/[m_\phi^2 \Gamma_\phi^2]$ and $\Gamma_\phi$ becomes proportional to the small coupling to the SM\@.

Finally, we display in gray the same theoretical constraints as in the di-muon case: at low $m_\phi$ annihilations to a pair of mediators dominate the relic density.

\subsection{Leptophilic Higgs}
\begin{figure*}[t]
\includegraphics[width=0.9\textwidth]{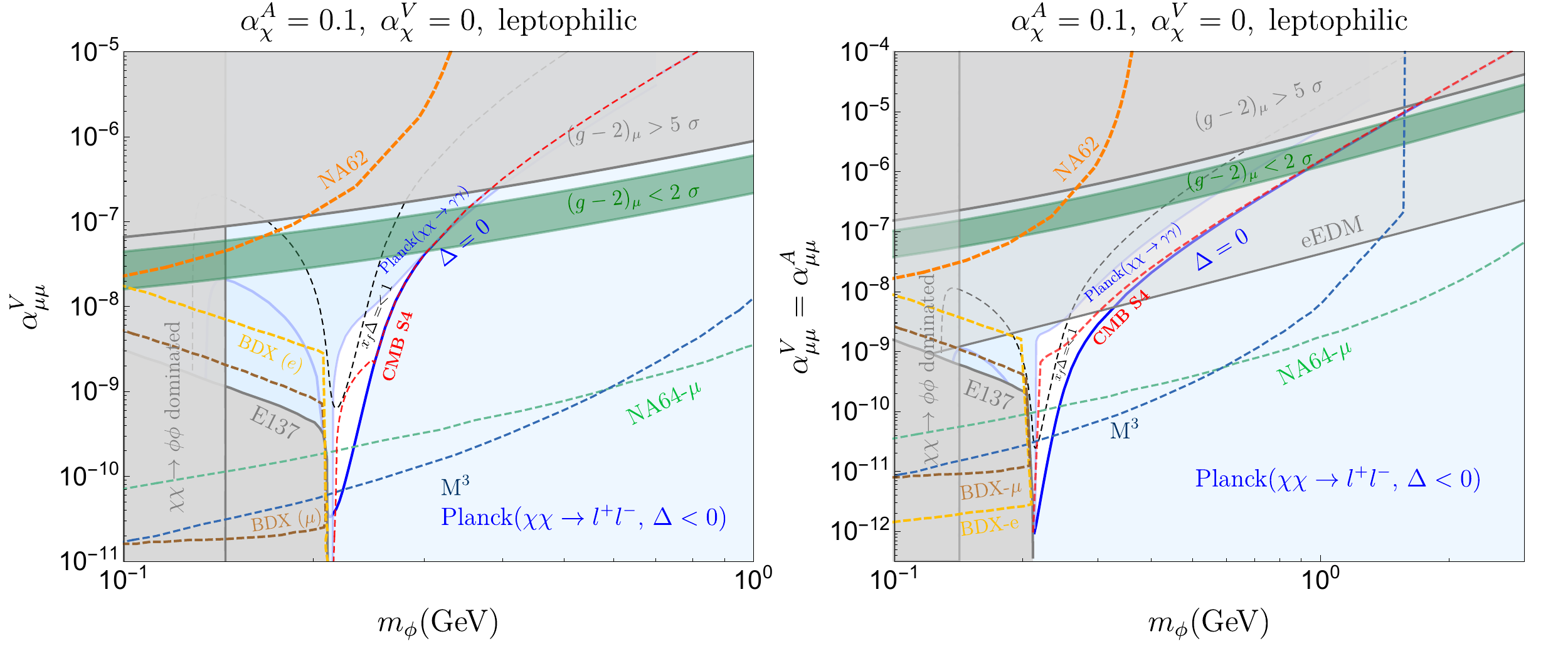}
\includegraphics[width=0.9\textwidth]{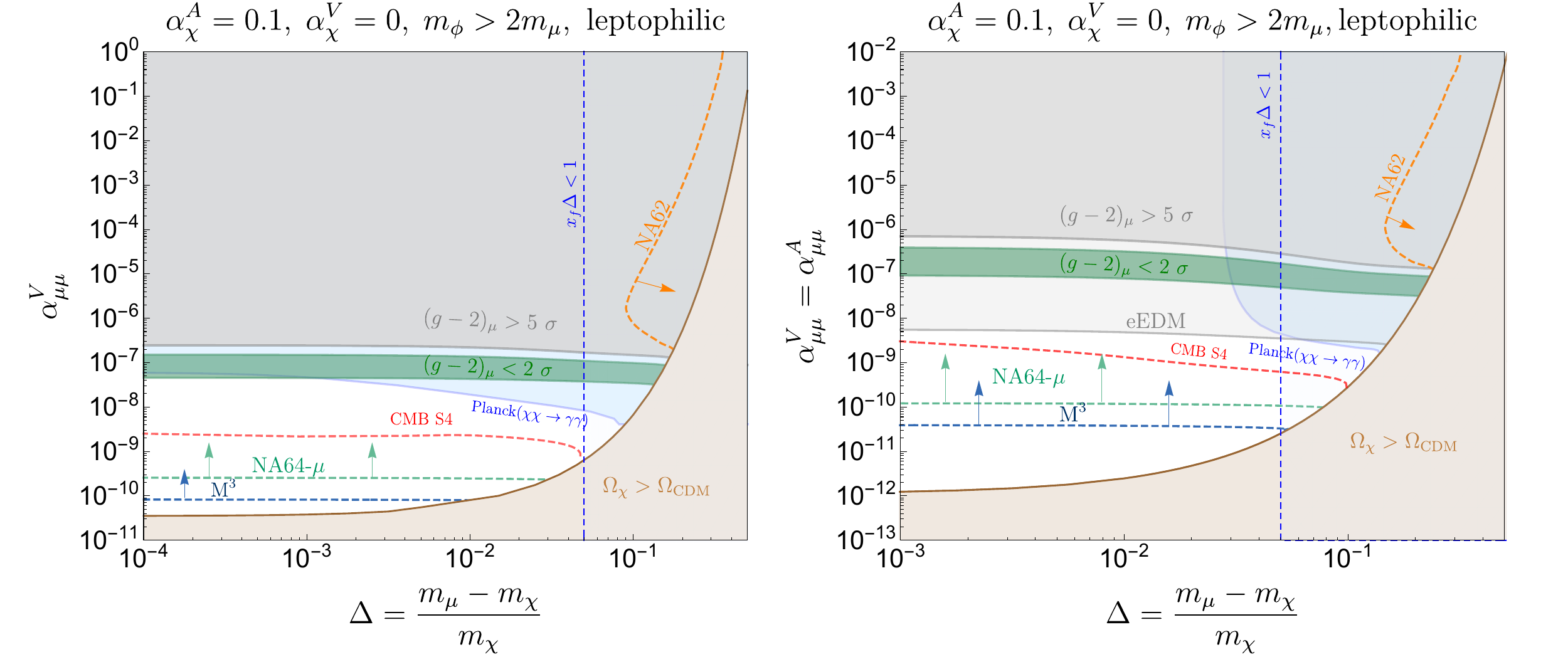}
\caption{
Theoretical and experimental constraints on Forbidden DM annihilations through a leptophilic light Higgs. We show constraints from the E137 electron beam-dump experiment~\cite{Marsicano:2018vin}, measurements of the muon anomalous magnetic moment~\cite{Bennett:2006fi, Hagiwara:2011af, Davier:2010nc}, the upper bound on the electric dipole moment of the electron~\cite{Andreev:2018ayy}, and Planck constraints on DM annihilations~\cite{Aghanim:2018eyx}. We include projections from CMB-S4 \cite{Abazajian:2016yjj} and the following future beam-dump experiments: BDX~\cite{Bondi:2017gul}, LDMX~\cite{Berlin:2018bsc, Kahn:2018cqs}, ${\rm M}^3$~\cite{Berlin:2018bsc, Kahn:2018cqs}, NA62~\cite{Krnjaic:2019rsv, NA62:2017rwk}, NA64-$\mu$~\cite{Chen:2018vkr, Gninenko:2014pea, Gninenko:2001hx}, and ${\rm M}^3$~\cite{Berlin:2018bsc, Kahn:2018cqs}.  In the upper (lower) row, we choose $\Delta$ ($m_\phi$) to match observed relic density in the white region. Ordinary annihilations into pairs of mediators set the relic density in the shaded gray region. In the brown area Forbidden DM has too large of an abundance.
}
\label{fig:lepto}
\end{figure*}

The single-channel annihilations discussed above essentially exhaust the description of the main phenomenological features of Forbidden DM annihilating to SM leptons. However it is interesting to consider explicitly scalar couplings to the SM proportional to lepton masses. The viable DM mass window is the same as in the di-muon case
\be
0.9 m_\mu \lesssim m_\chi \lesssim m_\mu\, .
\ee
For this range of DM masses there exists a range of DM couplings, shown in Fig.~\ref{fig:lepto}, for which kinematically forbidden $\chi \chi \to \mu^+ \mu^-$ annihilations dominate over the ordinary $\chi \chi \to e^+e^-$ annihilations. In this region of parameter space, the phenomenology is very similar to the pure $\chi \chi \to \mu^+ \mu^-$ case, due to the small electron mass in the SM, which suppresses its coupling: $\alpha_{ee} = (m_e^2/m_\mu^2)\alpha_{\mu\mu} $. DM masses close to $m_e$ are excluded by the $\Delta N_{\rm eff}$ constraint discussed in the $\chi \chi \to e^+e^-$ section. Masses closes to $m_\tau$ lead to ordinary annihilations into muon pairs dominating the Boltzmann equation. The constraints shown in Fig.~\ref{fig:lepto} and  the main theoretical features of the relic density calculation were already determined in Section~\ref{sec:mumu}\@. However the constraint from energy injection in the CMB from $\bar{\chi}\chi\rightarrow \gamma\gamma$ is more stringent than the muon-specific case, since all the three charged leptons contribute. As a consequence, almost all of the $(g-2)_\mu$ favored parameter space is ruled out. The other two novelties are a bound from the electron electric dipole moment induced by the axial coupling of the mediator to leptons \cite{Andreev:2018ayy, Lindner:2016bgg} and the projection from BDX-e~\cite{Marsicano:2018vin}, from the direct production of the leptophilic scalar $\phi$ through the  bremsstrahlung process off the primary electron beam, $e+N\rightarrow e+N+\phi$.

\section{Vector Mediators}\label{sec:vector}
Forbidden DM can not have large couplings to neutrinos, otherwise regular annihilations dominate its Boltzmann equations. The only viable choice involving a vector mediator within the SM lepton sector is to gauge right-handed muon or tau number. Annihilations into electrons are excluded by the same arguments discussed for scalar mediators in Section~\ref{sec:ele}\@. Gauging right-handed lepton number requires additional states.  We consider the addition of a pair of chiral leptons $E_{L, R}$ (anomalons) with hypercharge $\pm 1$ and muon(tau) number $0$ and $1$. One state cancels anomalies and the other generates a large enough mass for the pair. Furthermore, we assume the presence of a complex scalar $S$, charged under right-handed lepton number, that higgses the new gauge group. 
In practice we gauge a linear combination of right-handed muon (or tau) number and DM number. $\chi$ has charge $Q_\chi=1$,  the right-handed muon(tau) has charge $-Q$ and the scalar $S$ charge $+Q$. 

In addition to gauge interactions, at tree-level the leading interactions are
\be
- \mathcal{L} \supset y_i \frac{S}{M} \bar L_i H \ell_R + y_E S \bar E_L E_R + {\rm h.c.} \label{eq:vector}
\ee
As in the previous Section the first operator in Eq.~\eqref{eq:vector} can be generated by integrating out new vector-like leptons $X_{L,R}$
\be
-\mathcal{L}_{\rm UV}\supset \lambda_i \bar L_i H X_R + M \overline X_L X_R + \lambda S \overline X_L \ell_{R}+{\rm h.c.}
\ee
The LHC phenomenology of the new leptons $X$ was discussed in the previous Sections. To be consistent with experiment we take $M\gtrsim$~TeV\@. The chiral leptons $E_{L, R}$ are pair-produced at colliders through their coupling to hypercharge. In the absence of mass mixing with SM leptons they are stable on collider scales, which allows for essentially background-free searches~\cite{Chatrchyan:2013oca, Altmannshofer:2013zba, Aad:2015dha, Khachatryan:2016sfv, CMS:2016ybj, Aaboud:2018kbe, Jager:2018ecz}. Currently CMS sets the strongest constraint: $y_E \langle S \rangle \gtrsim 730$~GeV~\cite{CMS:2016ybj}. We also assume $\mathcal{O}(1)$ quartics and take also the scalar mass in the same ballpark $m_S \sim$~TeV\@. Given their masses, the new leptons $E_{L,R}, X_{L,R}$ and the new scalar $S$ do not appreciably affect the calculation of the relic density. 

\begin{figure*}[t]
\includegraphics[width=0.45\textwidth]{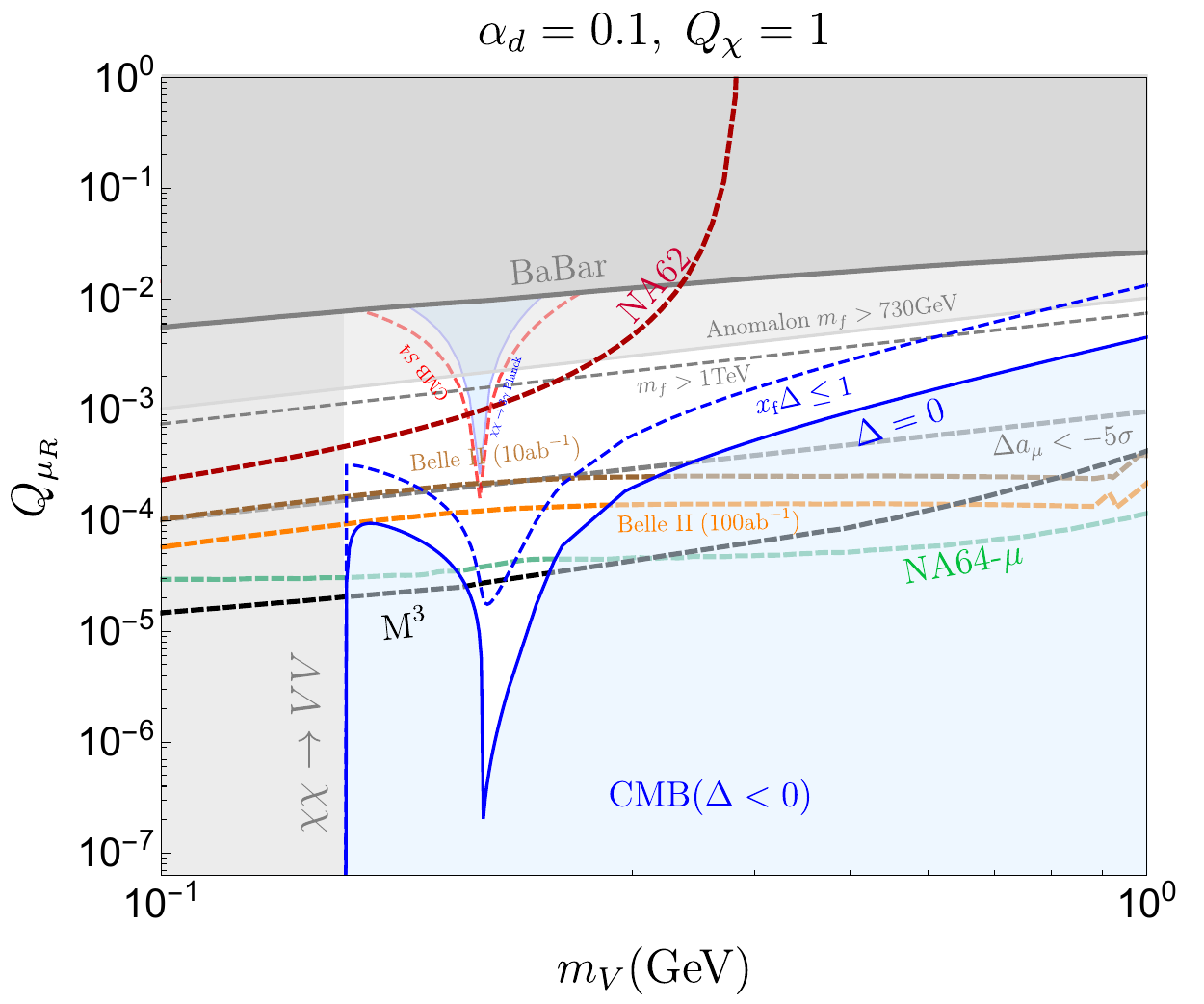}
\includegraphics[width=0.45\textwidth]{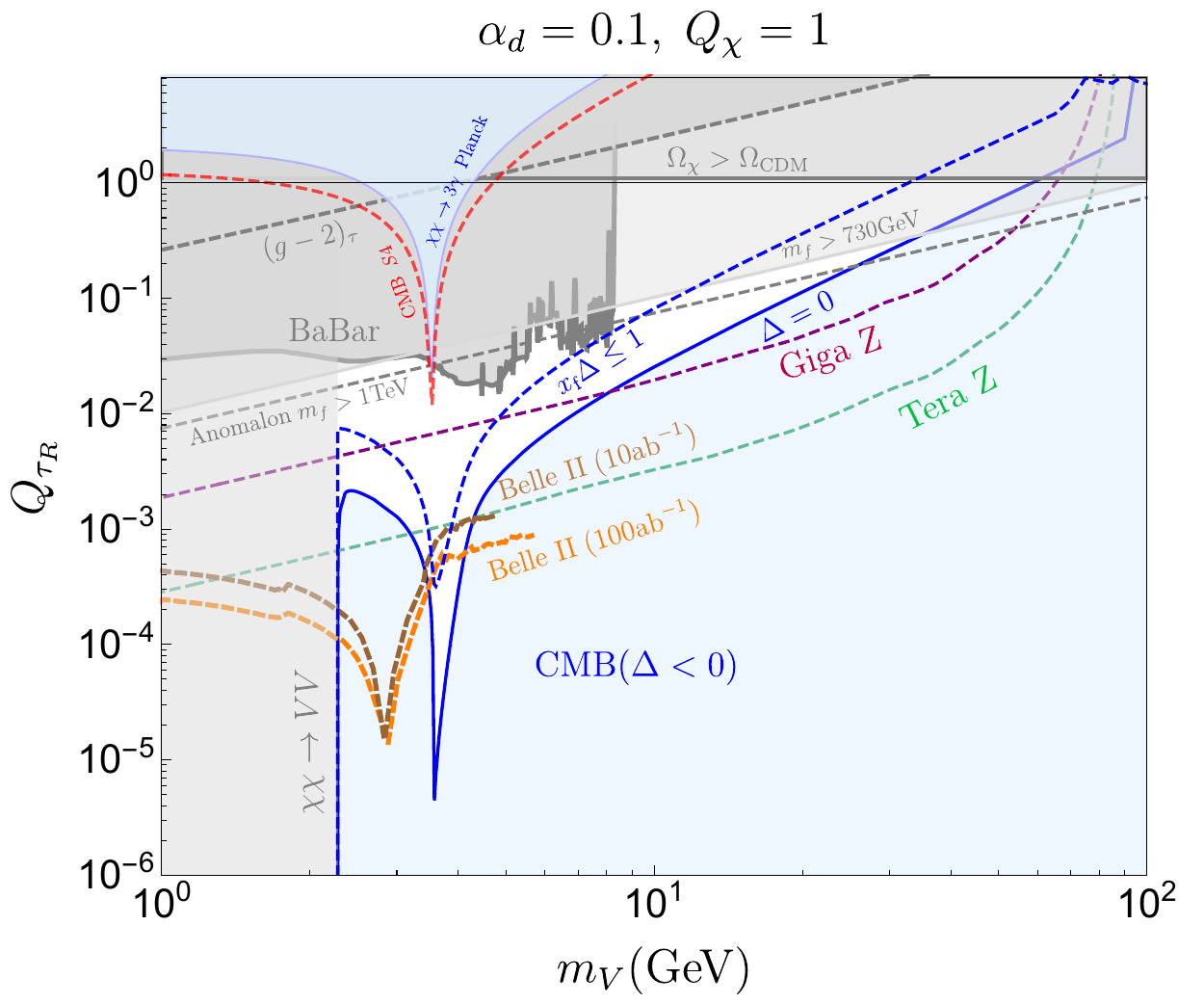}
\caption{
Theoretical and experimental constraints on Forbidden DM annihilations through a vector mediator coupled to right-handed muon number (left) and right handed tau number (right). We show constraints from measurements of the muon anomalous magnetic moment~\cite{Bennett:2006fi, Hagiwara:2011af, Davier:2010nc}, the tau anomalous magnetic moment~\cite{Beresford:2019gww}, BaBar searches for the vector mediator~\cite{Lees:2017lec}, and Planck constraints on DM annihilations~\cite{Aghanim:2018eyx}. We include projections from NA64-$\mu$~\cite{Chen:2018vkr, Gninenko:2014pea, Gninenko:2001hx}, Belle~II~\cite{Araki:2017wyg}, and future $Z$ factories~\cite{Liu:2017zdh}. In the white region $\Delta$ is chosen at each point to match the observed relic density. We shade in gray the region where ordinary annihilations to a pair of mediators set the relic density.}
\label{fig:vector}
\end{figure*}

From the point of view of DM phenomenology we have a mediator with a vector coupling to the dark sector, just like a dark photon, but a chiral coupling to the SM that involves only right-handed leptons. If coupled to muons, this vector mediator, with mass $m_V$, decreases $(g-2)_\mu$~\cite{Lindner:2016bgg, Kowalska:2017iqv}, 
\be
(g-2)_\mu= \frac{\lambda_V^2}{\pi} \int_0^1 dx \frac{\alpha_V^{\mu \mu} Q_1^+ (x)+\alpha_A^{\mu \mu} Q_1^- (x)}{(1-x)(1-\lambda_V^2 x)+\lambda_V^2 x}=\frac{\lambda_V^2 Q_{\mu_R}^2}{\pi} \int_0^1 dx \frac{2x(1-x)(3x-4)+4\lambda_V x^3}{(1-x)(1-\lambda_V^2 x)+\lambda_V^2 x}, \label{eq:g2V}
\ee
where $Q^{-}_1=2x(1-x)(x-4)+4\lambda_V^2 x^3$, $Q^{+}_1=2x^2(1-x)$, and $\lambda_V=m_\mu/m_V$. Therefore in a minimal model it is not possible to ease the tension between the SM prediction and observation. If  right-handed muon number is gauged,  a large fraction of the parameter space, in the absence of additional contributions, gives a deviation of $5\sigma$ or more with respect to the measured value of $(g-2)_\mu$, as shown in the left panel of Fig.~\ref{fig:vector} (gray dashed line). We do not discard this region as additional BSM contributions can in principle ease the tension which in the SM alone is at the $4.2\sigma$ level. We note that a mediator with a vector coupling to the SM gives a positive contribution to $(g-2)_\mu$, similar to the SM photon, easing the tension as for example in~\cite{Krnjaic:2019rsv}, but the axial part of our loop gives a larger negative contribution.

In Fig.~\ref{fig:vector} we show the phenomenology of Forbidden DM coupled through a vector mediator of mass $m_V$. The main features of the relic density are the same as in the scalar case. When $\Delta < 0$ and ordinary annihilations set the relic density, Planck measurements of the CMB exclude our light DM models. Forbidden annihilations ($\Delta >0$) in the white parameter space in Fig.~\ref{fig:vector} evade this constraint, except for a small region, shaded in blue, where one-loop annihilations $\bar \chi \chi \to 3 \gamma$ can affect CMB measurements~\cite{Aghanim:2018eyx}\footnote{Note that due to charge conjugation invariance, only the vector coupling to the SM contributes to this process~\cite{Laursen:1980ba,vanderBij:1988ac,Altarelli:1989wt,Glover:1993nv}. To compute the cross section for $\bar \chi \chi \to 3 \gamma$ relevant to energy injection into the CMB, we notice that it factors into $\bar{\chi}\chi\rightarrow A'^*$ and $A'^*\rightarrow 3\gamma$. The width for a massive photon decaying to three photons at one-loop order has been precisely calculated beyond the Euler-Heisenberg limit in~\cite{McDermott:2017qcg} from which we evaluate the off-shell dark photon decay width.}. As in the scalar case, we can have large Boltzmann suppressions $\Delta \sim \mathcal{O}(1)$ only for resonant annihilations $2m_\chi \sim m_V$. Furthermore, at low $m_V$ ordinary annihilations to pairs of mediators set the relic density. 

Laboratory constraints on our model include LHC searches for $E_{L,R}$~\cite{Altmannshofer:2013zba, Aad:2015dha, Khachatryan:2016sfv, CMS:2016ybj, Aaboud:2018kbe, Jager:2018ecz}, searches for monophoton events with missing energy via $e^+e^-\rightarrow \gamma A'$ followed by $A'\rightarrow$ invisible in 53 $\text{fb}^{-1}$ of $e^+e^-$ collision data collected with the BaBar detector at the PEP-II B-factory~\cite{Lees:2017lec}, and measurements of $(g-2)_\mu$~\cite{Bennett:2006fi, Hagiwara:2011af, Davier:2010nc} and $(g-2)_\tau$~\cite{Beresford:2019gww}. Searches for $\mu +N \to \mu+N+V$ at NA64-$\mu$~\cite{Chen:2018vkr,Gninenko:2014pea, Gninenko:2001hx} will be sensitive to most of the parameter space if right-handed muon number is gauged. In the scenario with gauged right-handed tau number, the non-decoupling effect of the anomalon provides sizable $Z\gamma A'$ and $\gamma^* \gamma A'$ vertices via the $U(1)_{\tau_R}U(1)^2_Y$ triangle diagram~\cite{Dror:2017nsg,Dror:2017ehi,Michaels:2020fzj,Ismail:2017fgq,Dedes:2012me}.  We show projections for searches for  monophotons at Belle~II~\cite{Araki:2017wyg} and $Z\rightarrow A'\gamma$ at future $Z$ factories. 

\section{Conclusions}\label{sec:conclusion}

Kinematically forbidden annihilations of DM were introduced as an exception to the standard WIMP relic abundance calculation by Ref.~\cite{Griest:1990kh}.  
Assuming a weak scale annihilation rate for DM, forbidden annihilations require  a small mass splitting between DM and its annihilation products ($\Delta \ll 1$)~\cite{Griest:1990kh}. 
 
More recently, forbidden annihilations were identified as a simple mechanism for light DM (with $m_\chi$ as low as the keV scale) that naturally satisfies stringent CMB constraints on energy injection at late times~\cite{DAgnolo:2015ujb}. This was achieved by considering larger mass splittings ($\Delta \sim \mathcal{O}(1)$) and annihilations into dark sector particles. The first application of annihilations into SM particles was presented in~\cite{Delgado:2016umt}, where DM predominantly couples to the top sector, leading to phenomenology not too dissimilar from a WIMP\@.

In this work we consider a new application of this mechanism that leads to qualitatively new predictions. We consider DM with a large coupling to the SM\@. Ordinarily, this leads to a relic abundance many orders of magnitude smaller than the observed one, with only trace amounts of DM present today. However, if the DM mass is in a window below the mass of its annihilation products, the relic density can have the observed value due to the thermal suppression of the annihilation rate. This leads to precise experimental targets for the DM mass that can be probed at future beam-dump experiments and high-luminosity colliders. In this work we have mapped annihilations into SM leptons finding three viable mass windows
\be
0.9 m_\mu \lesssim m_\chi \lesssim m_\mu\,, \quad  (m_\tau-m_\mu) \lesssim 2m_\chi \lesssim (m_\tau+m_\mu)\, , \quad 0.8 m_\tau \lesssim m_\chi \lesssim m_\tau\, .
\ee
In the first two windows this scenario can explain the discrepancy between the observed value of $(g-2)_\mu$ and the SM prediction~\cite{Aoyama:2020ynm, Bennett:2006fi, Hagiwara:2011af, Davier:2010nc, Grange:2015fou, Albahri:2021kmg, Albahri:2021ixb, Abi:2021gix}.
As noted in the introduction, the available parameter space for forbidden annihilations into the SM is reminiscent of WIMP DM in the MSSM\@. Much of the viable parameter space consists of special combinations of neutralinos, with special spectra, that can have the observed relic density consistent with experimental constraints~\cite{ArkaniHamed:2006mb}. Light DM, with a standard cosmological history and $s$-wave annihilations to SM particles, can have the observed abundance and satisfy CMB constraints~\cite{Padmanabhan:2005es, Aghanim:2018eyx} only for forbidden annihilations in narrow mass windows. This offers tantalizing targets for future experiments.  
For example, Forbidden DM annihilating into muons will be definitively probed by the next generation of fixed target experiments~\cite{Berlin:2018bsc, Kahn:2018cqs, Chen:2018vkr, Gninenko:2014pea, Gninenko:2001hx}. 

\vspace{.3cm}
\begin{acknowledgements}
{\em Acknowledgements---} 
We would like to thank A. Berlin, Q. Bonnefoy, X. Gan, C. Giovanetti, Y. Hochberg, E. Kuflik, M. Lisanti, H. Liu, and A. Rossia for  helpful discussions.  
JTR is supported by the NSF CAREER grant PHY-1554858 and NSF grant PHY-1915409.  JTR is also supported by an award from the Alexander von Humboldt Foundation and by the Deutsche Forschungsgemeinschaft (DFG, German Research Foundation) under Germany’s Excellence Strategy – EXC 2121 “Quantum Universe" – 390833306.
\end{acknowledgements}

\bibliographystyle{apsrev4-1}
\bibliography{biblio}

\end{document}